\documentstyle[eqsecnum,aps,preprint,tighten]{revtex}
\newcommand{\be}{\begin{equation}}
\newcommand{\ee}{\end{equation}}
\newcommand{\bea}{\begin{eqnarray}}
\newcommand{\eea}{\end{eqnarray}}

\begin{document}
\draft
\preprint{MPG-VT-UR 191/99}

\title{Topological gauge invariant variables in QCD}

\author{D.~Blaschke,~V.~N.~Pervushin\thanks{Permanent address:
Bogoliubov Laboratory of Theoretical Physics, Joint Institute for 
Nuclear
Research, 141980 Dubna, Russia} 
and 
G.~R\"opke}
\address{
Fachbereich Physik, Universit\"at Rostock, D-18051 Rostock, Germany
}


\maketitle

\begin {abstract}
We suggest that proper variables for the description of non-Abelian theories
are those gauge invariant ones which keep the invariance of the
winding number functional with respect to topologically nontrivial
(large) gauge transformations.
We present a model for these variables using the zero mode
of the Gauss law constraint and investigate their physical consequences
for hadron spectrum and confinement on the level of
the generating functional for two-color QCD.
\end{abstract}

\pacs{PACS number(s): 12.38.Aw}

\section{Introduction}
The identification of physical degrees of freedom in non-Abelian
gauge theories is a crucial  point for understanding the physical
phenomena hidden in their structure.
According to Dirac \cite{d}, the principle of local gauge invariance has 
to
be established not only on the level of the Lagrangian, but also on the
level of the variables used in the formulation of the gauge theory, 
since we
can observe only gauge-invariant quantities.
Dirac has obtained the unconstrained form of QED in terms of
gauge invariant variables for QED \cite{d} as functionals of the
initial gauge fields by explicitly resolving the Gauss law constraint.
The resulting unconstrained formulation of QED coincides with the one 
obtained in the Coulomb gauge~\cite{hp} with the physical phenomena of 
electrostatics, 'dressed' electrons, and two transverse photon degrees of 
freedom. The relativistic covariance of this formulation of QED has been 
proven by Zumino \cite{z} on the level of the algebra of generators of the 
Poincare group.

The Dirac definition of the gauge-invariant variables can be treated
also as a change of variables to construct the generating functional
of the Green functions in any gauges including the Lorentz invariant ones.
The invariance of the corresponding Green functions under a change of 
variables (which generates the Ward-Taylor-Slavnov identities \cite{fs}) 
is  guaranteed by the Dirac factors in source terms, which
restore the Coulomb gauge Feynman rules in any Lorentz invariant gauge.
So, the Coulomb interaction and electrostatics are consequences
of the identification of the physical degrees of freedom which correspond 
to an explicit solution of the Gauss law, but not primarily to the choice of 
the gauge. For example, 
if one would omit the Dirac factors in for the source terms in relativistically
invariant Lorentz gauge formulations of QED, one would get
the Wick-Cutkosky bound states formed by gauge propagators with
light-cone singularities with a spectrum different
\footnote{One of the authors (V.P.) thanks W. Kummer who pointed out that in 
Ref. \cite{kum} the difference between the Coulomb atom and
the Wick-Cutkosky bound states in QED has been demonstrated.}
from the observed one which corresponds to the instantaneous Coulomb 
interaction.
Thus, the Dirac variables in QED are gauge-invariant, Lorentz covariant,
and bear direct relation to measurable quantities.

The attempts to obtain the non-Abelian generalization of the
Dirac gauge invariant variables
\cite{vp1,vp2,kp,ml,hal}  meet the difficulties of
the nontrivial topological structure of gauge transformations which are
classified as 'small' and 'large' ones in accordance with their
homotopy group, e.g. $\pi_{(3)}(SU(2))=Z$ \cite{jac,jac1} for the example of
two-color QCD considered below.
The quantization of the fermion sector of the theory leads to
anomalies \cite{ABJ} and to the topological 'winding number'
functional of the gauge fields which plays an 
important role in the description of physical phenomena
connected with the $U_A(1)$ anomaly \cite{w1,1,2,3,bl}.
It is, however, not invariant with respect to large 
gauge transformations and the problem arises: Is it possible
to construct 'topological' gauge invariant variables
which leave the winding number invariant under large gauge transformations?

In the  present paper, we study the physical consequences of the nontrivial
topology of non-Abelian gauge fields using topological Dirac variables. 
The resulting representation of the QCD generating functional is defined by 
the zero modes of the Gauss law constraint \cite{vp1,vp2,kp} in the class of
functions of large gauge transformations, which disappear at spatial
infinity but have nonzero surface integrals.

The paper is organized as follows. 
Section 2 reviews the introduction of ordinary Dirac variables for non-Abelian 
theories using the example of the $SU_c(2)$ Yang-Mills theory.
In Section 3, we give the definition of topological Dirac variables which are
constructed by including a zero mode of the Gauss law constraint. 
We derive an equation for the topological invariance of non-perturbative fields
in the class of functions corresponding to large gauge transformations.
In Section 4, we show that the Wu-Yang monopole is a solution of this equation 
and use it to obtain the electric field of the $\theta$-vacuum, the 
instantaneous interaction for QCD, and equations for quasiparticle excitations.
In Section 5, the generating functional for Green functions is constructed
and a discussion of the problems of hadronization and confinement is given.
The conclusions are presented in Section 6.

\section{Dirac variables in non-Abelian theories}
\label{sec:dirvar}

We consider two-color QCD, i.e. $SU_c(2)$ Yang-Mills theory coupled to 
fermionic fields (quarks), with the action functional
\be \label{u}
W[A,\Psi]=\int dt\int\limits_{V} d^3x
\left(-\frac{1}{4} G^2+ \bar\psi[i\gamma^\mu(\partial _\mu+{\hat A_\mu})
-m]\psi\right)~,
\ee
where ${\hat A}=g\frac{\tau^a A^a}{2i}~$,
\be \label{g2}
-\frac{1}{4} G^2=\frac{1}{2}\sum_{i,a}({E^a_i}^2- {B_i^a}^2)~,
\ee
and the conventional notations for for the covariant derivative
$D^{ab}_i(A):=\delta^{ab}\partial_i + g\epsilon^{acb} A_i^c$ as well as for the
non-Abelian electric and magnetic fields
\be \label{v}
E_i^a = \partial_0 A^a_i - D_i^{ab}(A)A_0^b~,
\ee
\be
B_i^a=\epsilon_{ijk}\left(\partial_jA_k^a+
\frac g 2\epsilon^{abc}A^b_jA_k^c\right)~,
\ee
respectively, have been used.

The action (\ref{u})  is invariant with respect to arbitrary gauge
transformations  $u(t,x)$
\be \label{gauge1}
\psi^u := u(t,x)\psi~,
\ee
\be \label{gauge2}
{\hat A}_{i}^u := u(t,x)\left({\hat A}_{i} + \partial_i
\right)u^{-1}(t,x)~.
\ee
In order to eliminate unphysical degrees of freedom, we introduce the
non-Abelian generalization of the Dirac variables~\cite{d}
according to \cite{vp1}
\be \label{y1}
\psi^D[\psi,A]:=U[A]\psi~,
\ee
\be \label{y2}
{\hat A}_{\mu}^D[A] := U[A]\left({\hat A}_{\mu} + \partial_{\mu}
\right)U^{-1}[A]~,
\ee
where  $U[A]$ is a solution of the linear differential equation
\be \label{DU}
U[A]\left( \hat{a}_0[A] + \partial_0\right) U^{-1}[A]= 0~,
\ee
and $a_0[A]$ is a solution of the Gauss equation without the
fermion source term
\be \label{GU}
[D^2(A)]^{bc}a^c_0= [D_k(A)]^{bc} \partial_0 A_k^c~.
\ee
By construction, the matrix $U[A]$
has  the transformation property \cite{vp1}
\be
U[A^u]=U[A] u^{-1}~,
\ee
so that the Dirac variables $A_i^D$
are  gauge-invariant functionals of the initial gauge potentials
\be  \label{ga}
\hat A^D[A^u]=\hat A^D[A]~, \quad \quad 
\psi^D[\psi^u,A^u]=\psi^D[\psi,A]~,
\ee
which satisfy the identity
\be \label{b1}
D_i^{ab}(A^D){\partial_0 (A_i^D)^b}\equiv 0~.
\ee
The solution of the system
of linear differential equations
(\ref{DU}), (\ref{GU})
can be written in the form
of the time ordered exponential
\be
U(t, \vec{x}; t_0)= v(\vec{x})
T\mbox{exp}\left(\int^t_{t_0}dt{\hat a}_0(t, \vec{x})\right)~.
\ee
The gauge-invariant Dirac variables $A^D$ as solutions of the 
differential
equation (\ref{DU}) are defined for
the initial values
\be  \label{id}
U(t, \vec{x})\mid_{t=t_0}= v(\vec{x})~.
\ee
These values define
the remaining group of stationary gauge transformations
of the Dirac variables.
The group of the stationary gauge transformations $v(\vec{x})$ in the
three-dimensional coordinate space is topologically nontrivial and represents
the group of three-dimensional paths lying  on the three-dimensional space of
the $SU_c(2)$-manifold with the homotopy group $\pi_{(3)}(SU_c(2))=Z$.
The whole group of the stationary gauge transformations is split into
topological classes marked by the degree of the map
(i.e. the integer number $n$) which counts
how many times a three-dimensional path turns around the $SU(2)$-manifold
when the coordinate $x_i$ runs over the space where it is defined.
The stationary
transformations $v^n(\vec{x})$ with $n=0$ are called the small ones;
and those with $n \neq 0$
\be \label{gnl}
{\hat A}^{(n)}=v^{(n)}(\vec{x})({\hat A}+\partial){v^{(n)}(\vec{x})}^{-1}~,
\ee
the large ones.

Quantization of the fermion sector of the theory leads to the
well-known Adler-Bell-Jackiw anomaly \cite{ABJ}
\be
W_{\rm anomal}[A;j,\eta,\bar \eta]=
C_{\eta}\int dt~\bar \eta I_c \gamma_5 \eta~\frac{d}{dt}X[A]~,
\ee
where $\eta,\bar \eta$ are fermion sources, $C_\eta$ is a constant and 
$I_c$ is the unit matrix in color space.
\be \label{e1}
 X[A^D]=-\frac {1}{8\pi^2}\int\limits_V d^3x
\epsilon^{ijk}Tr \left[{\hat A^D}_i\partial_j{\hat A^D}_k -
 \frac 2 3 {\hat A^D}_i{\hat A^D}_j{\hat A^D}_k\right]
\ee
is the topological winding number functional of the gauge fields.
This functional plays an important role for the description of
observable phenomena in hadronic physics connected with the violation
of the axial $U_A(1)$ symmetry and, in particular, with the occurrence of the
$\eta-\eta'$ mass difference \cite{w1,1,2,3,bl}.
The functional $X[A]$ is not invariant with respect to large gauge
transformations (\ref{gnl})
\cite{jac} (see defs. (3.33), (3.36) in \cite{jac1}):
\be \label{gx}
X[A^{(n)}]=X[A]+{\cal N}[b,n]~,
\ee
where
\be \label{gn}
{\cal N}[b,n]
=\frac {1}{8\pi^2}\int d^3x ~\epsilon^{ijk}~ Tr[\partial_i({\hat b}_j L^n_k)
- \frac 1 3 (L^n_iL^n_jL^n_k)]~.
\ee
$L^n_k=v^n\partial_k(v^n)^{-1}$ is a pure gauge field and
$\hat b_i(\vec x)$ is an asymptotics of the field $\hat A^D_i(t,\vec x)$
at the spatial infinity
\be \label{b}
  \lim_{x \rightarrow \infty} A^D(t,x) =b(x).
\ee
Both the large transformations and 'asymptotic' field $b(x)$
are given in the electrostatic-type class of the functions
which disappear at spatial infinity but have nonvanishing
surface integrals in Eq. (\ref{gn}).

In the context of the Dirac invariant description of the quantum theory
the question appears: Is it possible
to construct 'topological' non-Abelian variables as
functionals of the Dirac ones $A^T[A^D]$, so that the winding number
~(\ref{e1}) becomes invariant also with respect to
the large gauge transformations, i.e.
\be \label{gdn1}
 X[A^T[{A^D}^{(n)}]] = X[A^T[A^D]]
\ee
for $n \not =0$ ? The answer to this question is the subject of the present 
paper and will be given in the next Section.

\section{Topologically invariant Dirac variables} \label{sec:topdir}

\subsection{Zero mode of the Gauss constraint}

Our idea is to construct such topological variables $A^T$
for which the winding number (\ref{gdn1}) converts into
a zero mode of the Gauss constraint \cite{vp1,vp2,kp,n}.
The latter, in terms of the Dirac variables~(\ref{y1}),~(\ref{y2}),
has the form
\be\label{gaussd}
(D^2(A^D))^{ac} ({ A^D_0})^c = (j^D_0)^a~,
\ee
where $j_{\mu}^a= \frac{g}{2}\bar\psi\tau^a\psi$ is the fermionic current.
A general solution of this inhomogeneous equation is
a sum of the solution $\Phi^c$  of the homogeneous equation
\be\label{homl}
\left(D^2(A)\right)^{ac} \Phi^c = 0~,
\ee
i.e. a zero mode of the Gauss law constraint, and a partial solution 
${\bar A}^D_0$ of the inhomogeneous one~(\ref{gaussd})
\be \label{genl}
A^D_0 = -\Phi + {\bar A}^D_0 .
\ee
The electric field strength
\be \label{ed}
E_k^a = \partial_0 (A^D_0)^a + D_k^{ac}(A^D) (A_0^D)^c
\ee
can also be decomposed into a zero  mode part and a perturbative part
${\bar E}_k^c$~,
\be \label{edg}
E_k^c = D^{cb}_k(A^D)\Phi^b + {\bar E}_k^c~.
\ee
Then, the initial Yang-Mills action in (\ref{u}) is a sum
of global (G) and local (L) parts
\be
W_{YM} = W_G+ W_L~.
\ee
The local part coincides with the gauge field sector of the initial 
action~(\ref{u}) with the fields $A_0$ replaced by ${\bar A_0}$, and 
the global one is
\be \label{gp}
 W_G= \int dt ~(I_E + I_\Phi) ~,
\ee
where
\bea \label{ie}
I_E &=& \int d^3 x ~[{\bar E}_iD_i(A) \Phi + j_0\Phi] ~,\\
\label{IPhi}
I_\Phi &=& \frac{1}{2}\int d^3 x ~(D_i(A) \Phi)^a (D_i(A) \Phi)^a ~.
\eea

To construct the generating functional in quantum theory~\cite{al},
we need only the constraint-shell
action, where the (global) zero-mode part converts into a sum of
two surface integrals~\cite{kp}
\bea \label{IE}
I_E &\equiv& \int d^3 x ~\partial_i ({\bar E}_i \Phi)\;=\;
\oint ds_i ({\bar E}_i \Phi)
\Bigl\vert_{\mid {\vec  x} \mid~\to~\infty }\,
\\
\label{IP}
I_\Phi&\equiv& \frac{1}{4} \int d^3 x ~\Delta (\Phi^a)^2~.
\eea
Thus, we can restrict ourselves to the 'electrostatics'  class of
functions of topologically nontrivial gauge transformations,
in the region of the spatial infinity.
For symplicity, we suppose that the quasiparticle part
disappears, $I_E = 0$, and the Dirac field
$A^D$ at spatial infinity converts into a stationary one
($b_i(\vec x)$, Eq.~(\ref{b})).
In this case, the zero mode field factorizes into
\be
\label{factorization}
\Phi^a(t,\vec x)\,\Bigl\vert_{\mid {\vec  x} \mid~ \to~ \infty }\;=
\;\dot N_0(t)\Phi_0^a (\vec x)~,
\ee
where $\Phi_0^a(\vec x)$ satisfies the equation
\be \label{homd}
(D^2(b))^{ca} {\Phi_0}^a = 0~.
\ee
$N_0$ is a zero mode
with the dynamics of a free rotator \cite{vp1} defined by the Lagrangian
~(\ref{IP})
\begin{equation}
\label{ktg}
I_{\Phi}=\frac{M_N}{2}{\dot N}_0^2 ;~~~
M_N=\frac 1 2\int\limits_{V}d^3x ~\triangle (\Phi^a_0)^2~.
\end{equation}

\subsection{The topological Dirac variables}

We define the topological variables 
\bea \label{gdv}
{\hat A}^T[A] &=&{\bar U}[\Phi^D]({\hat A}^D + \partial){\bar U}^{-1}[\Phi^D]~,
\\
\label{bu}
{\bar U}[\Phi^D]&=&
T \exp \left(\int^t_{t_0}dt{\hat \Phi^D}(t, \vec{x})\right),
\eea
so that the topological functional (\ref{e1}) in terms of
these variables depends only on the zero mode $N_0$:
\be  \label{top}
X[A^T]= X[A^D]+{\cal N}(b,N_0-X[A^D])=N_0~,
\ee
where
\be \label{314}
{\cal N}(b,N)
=\frac {1}{8\pi^2}\int d^3x \epsilon_{ijk} Tr[\partial_i({\hat b}_j
L^N_k) - \frac 1 3 (L^N_iL^N_jL^N_k)]~,
\ee
and $L^N_i$ is the pure gauge field
\bea
L^N_k&=&U(N)\partial_k(U(N))^{-1}~,\\
\label{limit}
U(N)&=&\lim_{\mid {\vec  x} \mid ~\to~ \infty} {\bar U}[\Phi^D] =
\exp[N(t) {\hat \Phi}_0({\vec x})]~.
\eea
In order to fulfill the constraint (\ref{top}), it is sufficient to find
the magnetostatic field  $\hat b_i(\vec x)$ for which
\be   \label{cond}
{\cal N}(b,N)=N~.
\ee
This self-consistency condition for the invariance of the topological
functional~(\ref{314})
is a necessary and sufficient condition for the topological invariance 
of
the Dirac variables~(\ref{gdv}),~(\ref{bu}) constructed above.

Let us solve the constraint~(\ref{cond}) for the class of functions
of topologically nontrivial gauge transformations \cite{jac,jac1}
with the boundary conditions $f(0)=0,~~ f(\infty)=1$
supposing that the corresponding zero mode solution
${\hat \Phi}_0({\vec x})$ has the form
\be  \label{cond2}
{\hat \Phi}_0({\vec x}) = -in^a\tau^a \pi f(r);~~~r={\mid {\vec  x}} ~.
\mid
\ee
In this case, the second part of ${\cal N}(b,N)$ in Eq. (\ref{314}) is
equal to \cite{jac1}
\be  \label{sp}
(N-\frac{\sin(2\pi N)}{2\pi})~.
\ee
Then, the first part should be equal to
$$(\frac{\sin(2\pi N)}{2\pi})~,$$
in order to fulfill the condition~(\ref{cond}) ${\cal N}(b,N)=N$.
One can be show that this compensation is fulfilled for an asymptotic
field $b(\vec{x})$ which has the form of the Wu-Yang monopole \cite{wy,fn},
i.e.
\be  \label{s}
b_i^a=\frac 1 g \epsilon_{iak}\frac{n_k(\Omega)}{ r}~,~~~~~~
n_k(\Omega)=\frac{x^l\Omega_{lk}}{r}~,
\ee
where $\Omega^{lk}$ is an orthogonal matrix in  color space:
$n_k(\Omega)n^k(\Omega)=1$~.

The Wu-Yang monopole and similar solutions of the classical equations
are also present in $SU(3)$ theory,
if we choose the minimal subgroup $SU(2)$ of $SU_c(3)$ (this means that the
fundamental representation of $SU_c(3)$ is an irreducible one  of
this subgroup).

For example, the role of matrices $ \tau_1,\tau_2,\tau_3 $
of the minimal subgroup $SU(2)$,
in $SU_c(3)$ theory,
is played by $(\lambda_2,\lambda_5,\lambda_7)$.
In this case,
\be  \label{rp}
 \hat =
b_i=g\frac{b_i^1\lambda^2+b_i^2\lambda^5+b_i^3\lambda^7}{2i};~~~~~
b_i^a=\frac{\epsilon^{aik}n^k}{gr}.
\ee
Thus, we can reproduce the construction of the topological Dirac variables
also in the $SU_c(3)$ non-Abelian theory.

\section{Physical consequences}

\subsection{Non-perturbative asymptotic field}

In order to illustrate physical consequences of the topological
Dirac variables, we consider the Wu-Yang monopole more in detail.

The Wu-Yang monopole satisfies the classical equation  everywhere
except for the singularity at $r=0$ with the corresponding magnetic field
\be  \label{sb}
B_i^a(b)=\frac{n^a n^i}{gr^2}~.
\ee
Following
Wu and Yang \cite{wy}, we consider the whole finite space volume,
excluding  an $\epsilon$-region around the singular point.
The size of this small region is chosen so that
the vacuum energy of the monopole  solution (\ref{s})
\be  \label{eb}
\frac 1 2 \int d^3x(B_i^a(b))^2=\frac
1{2\alpha_s}\int\limits_{\epsilon}^{R}\frac {dr}{r^2}\simeq\frac 1 2 
\frac 1
{\alpha_s\epsilon} +O[\frac 1 R ]~,~~~~~~~\alpha_s=\frac{g^2}{4\pi}
\ee
is removed by a finite counter-term in the Lagrangian,
  $$\bar {\cal L}={\cal L}-\frac{\mu^4}{2\alpha_s} ~.$$
The choice $\mu^4=(\epsilon V)^{-1}$ leads to volume-independent results.
The parameter $\epsilon$ of the ultraviolet cutoff 
is  proportional to  $V^{-1}$ and disappears in the infinite volume limit
 \be  \label{v1}
(\epsilon)^{-1}=V{\mu}^4~,
\ee
where the parameter $\mu$ has the dimension of energy
and determines the average density of the asymptotic field $b$
\be  \label{cd}
\frac{1}{2V}\int d^3xG^2(b)=\frac{<\alpha_s G^2(b)>}{2}={\mu}^4~.
\ee

\subsection{$\theta$-vacuum}

For the given
regularization there is a solution of the zero mode equation (\ref{homl})
in the form of Eq. (\ref{cond2})  
\be  \label{sz}
\Phi^a=\frac{1}{g}n^a2\pi f(r);~~~~~~~ f(r)=1-\frac{\epsilon}{r}~,
\ee
with the function $f(r)$ which fulfills the boundary conditions 
$f(\epsilon)=0$ and $f(\infty)=1$, in the considered region of space. 

In this case, the functional~ (\ref{314}) does not depend
on the parameter of regularization and can be calculated exactly.
As we have seen above, it coincides with the zero mode variable,
${\cal N}(b,N)=N$.

The kinetic term $I_{\Phi}$ is equal to
\bea  \label{kt}
I_{\Phi}&=&\frac{M_N}{2}{\dot N}_0^2~;\\
M_N&=&\frac 1 2 \int\limits_{V}d^3x ~\triangle
(\Phi^a_0)^2=
\frac{4\pi^2}{\alpha_s}\int\limits_{\epsilon}^{R}dr
~\frac{d}{dr}(r^2\frac{d}{dr}f)=
\frac{4\pi^2}{\alpha_s}(\mu^{4}V)^{-1}~,\nonumber
\eea
where we took into account Eq.~(\ref{v1}).
The zero mode $N_0$ is given in the physical region
$0 \le N_0 \le 1$, where the endpoints $N_0=0, N_0=1$  are physically 
equivalent, so that the phase space of the physical variables has the 
topology of a cylinder \cite{jac,vp2}. 
The Lagrangian of the global motion (\ref{kt}) describes a free rotator with
the momentum spectrum \cite{vp1}
$$
P_0= {\dot N_0}{M_N}= (2\pi k + \theta)~, ~~~~k=0,1,2,... ,
$$
which follows from the constraint
on the wave function
$$\Psi(N_0+1)=\exp(i\theta)\Psi(N_0).
$$
The corresponding vacuum part of the electric field in (\ref{ed}) does not
contain the regularization parameter and reads
\be\label{se}
E_i^a=\dot N_0 ~(D_i(b) \Phi)^a=(k+\frac{\theta}{2\pi}) \alpha_s B_i^a~,
\ee
where $B_i^a$ is the vacuum magnetic field (\ref{sb}).
The vacuum state with the minimal wave number $k=0$
corresponds to a nonzero electric field
\be\label{sem}
(E_i^a)_{\rm min}=\frac{\theta \alpha_s B_i^a}{2\pi}~,
\ee
i.e. a 'persistent field motion' around the 'cylinder'.
We have obtained a field theoretical analogy of the Josephson effect: a 
circular current without sources.
Coleman \cite{col} was the first who guessed an effect similar to (\ref{sem})
in $QED_{1+1}$, but from a classical point of view. The quantum
treatment of this effect was discussed in Refs. \cite{vp2,jackiw}.

This 'duality'  of the vacuum in the Minkowski space (\ref{se}) shows
that there can be a value of the superfluid momentum $P_0=2\pi/\alpha_s$
for which the vacuum Lagrangian vanishes
$$
\alpha_s=(k+\frac{\theta}{2\pi})^{-1}~~\rightarrow~~ 
G^2=2(B^2-E^2)=0
$$
without any counterterm  discussed before.

\subsection{Non-Abelian generalization of the Coulomb potential}
We can also calculate the instantaneous interaction which corresponds to 
the Green function of the gauge field when perturbations of the vacuum are 
neglected, see (\ref{edg}). 
In the presence of the Wu-Yang monopole the Green function satisfies the 
equation
$$
(D^2)^{ab}({\vec x})G^{bc}({\vec x},{\vec y})=
\delta^{ac} \delta^3({\vec x}-{\vec y})~,
$$
where
$$
(D^2)^{ab}({\vec x})= \delta^{ab}\Delta-
\frac{n^an^b+\delta^{ab}}{r^2}+
2(\frac{n_a}{r}\partial_b-\frac{n_b}{r}\partial_a)~,
$$
and $n_a(x)={x_a}/{r};~ r=|{\vec x}|~$.
Let us decompose $G^{ab}$ into a complete set of orthogonal vectors in
color space
$$
G^{ab}({\vec x},{\vec y})=  [n^a(x)n^b(y)V_0(z) + 
\sum\limits_{\alpha=1,2}
e^a_{\alpha}(x) e^b_{\alpha}(y)V_1(z)];~~~(z=|{\vec x}-{\vec y}|)~.
$$
Substituting the latter into the first equation, we get
$$
\frac{d^2}{dz^2}V_n+ \frac{2}{z} \frac{d}{dz}V_n- \frac{n}{z^2}V_n=0~.
$$
The general solution for the last equation is
$$
 V_n(z)= d_n z^{l_1^n} + c_n z^{l_2^n};~~~(n = 0,~1)~,
$$
where $d_n$, $c_n$ are constants, and
${l_1^n},~{l_2^n}$ can be found as  roots of the equation
$(l^n)^2 + l^n -n =0$; i.e.
$$
{l_1^n} =-\frac{1+\sqrt{1+4n}}{2};~~~
{l_2^n} =\frac{-1+\sqrt{1+4n}}{2}~.
$$
It is easy to see that for $n=0$ we get the Coulomb potential 
$d_0=-1/4\pi$,
and for $n=1$  the 'golden section' potential with
$$
{l_1^1} =-\frac{1+\sqrt{5}}{2}\approx - 1.618~
 ;~~~{l_2^1} =\frac{-1+\sqrt{5}}{2}\approx 0.618~,
$$
which can be used as the potential for the 'hadronization' of two-color QCD.

\subsection{Quasiparticle excitations}

The presence of the Wu-Yang monopole $\hat{b}_i(\vec{x})$ does not change
the qualitative character of
the excitation spectrum of the perturbation theory, it only mixes the
color and spin-orbital quantum numbers.
To demonstrate this fact, let us consider the fundamental representation
with the equation for a fermion in the Wu-Yang  monopole
\be \label{qpsi}
i\gamma_0 \partial_0 \psi + \gamma_j [i\partial_j \psi+ \frac{1}{2r}
\tau_a \epsilon^{ajl}n_l \psi] = 0~.
\ee
In the two-component form of the fermion field,  $(\psi_+,~\psi_-)$,
each component is a $2\times 2$ - matrix which can
be decomposed into
a scalar $s_{\pm}$ and a vector $v_{\pm}^j$~,
\be \label{qsv}
 \psi_{\pm}^{\alpha,\beta} =s_{\pm} \delta^{\alpha,\beta}
+ v_{\pm}^j\tau_j^{\alpha,\beta}~.
\ee
Finally, we obtain the following set of equations for
the scalar $s_{\pm}$ and vector $v_{\pm}^j$ amplitudes 
\bea
(\mp q_o+m) s_{\mp} {\mp}i(\partial_a + \frac{n_a}{r}) v^a_{\pm}&=&0~,\\
(\mp q_o+m) v^a_{\mp} {\mp}i(\partial^a - \frac{n^a}{r}) s_{\pm}
-i\epsilon^{jab}\partial_jv_{\pm}^b &=&0~,
\eea
where $q_0$ is an eigenvalue of the $i\partial_0$ - operator.
These equations are solved by decomposing the functions $(s,v)$ w.r.t.
orbital momenta. As a result, we obtain a continuous spectrum of
the conventional perturbation theory with a mixing of color and spin
and the wave function asymptotics $O(r^{-n})~, n > 1$.

\section{Generating functional}

\subsection{Problems of the covariant Coulomb gauge}

Let us consider  differences between the topological Dirac variables
$A^T, \psi^T$ and the conventional variables~\cite{sch,al} for
covariant Coulomb gauge
\be
D_k(b) A_k =0
\ee
on the level of the generating functional of the unitary perturbation
theory treating the asymptotic field $b({\vec x})$ as a nonperturbative
background.
For comparison, it is worth to repeat the construction of the
generating functional given in Ref. \cite{al}
for the case of an external field (here, the Wu-Yang monopole)
\begin{eqnarray}\label{sch}
Z_D[J,\eta,\bar\eta]&=&\int [d A_i][d E_i] [d \bar\psi] [d \psi]
\Pi_x \delta(D_k E_k) \delta(D_k A_k)\nonumber\\
&&\exp\left\{i\int d^4x \left[E_k\cdot \dot A_k -\frac{1}{2}E_k^2
 -\frac{1}{2}B_k^2 - \frac{1}{2}(D_k f)^2 - A_k \cdot J_k^c
 - \bar \psi \eta^c - \bar \eta^c \psi \right]\right\} ~~.
\end{eqnarray}
This construction is based on an explicit solution of the Gauss law 
constraint obtained by decomposing the electric field components of the field
strength tensor $G_{0i}$ into transverse $E_i$ and longitudinal 
$G^L_{0i}=-D_if$ parts,
\be \label{long}
 G_{0i}=E_i -D_i(b)f;~~~ D_k(b) E_k =0 ~.
\ee
Here the function $f$ can be determined from the equation \cite{sch,al}
\begin{equation}
\left((D^2)^{ab} + g\epsilon^{adc}A^d D^{cb}  \right) f_b  = j^b_{{\rm tot},0};
~~~~j^b_{{\rm tot},0}=g\epsilon^{abc}A_i^a E_i^{c}+j^b_0~,
\end{equation}
where $D\equiv D(b)$.

As we have seen above, the Wu-Yang monopole
leads to the instantaneous  potential
of hadronization which can form mesonic bound states.
In order to describe this bound state, we can use the bilocal representation 
of the generating functional in the meson channel
with the external fields $(b+A)$ \cite{pre,vp3}
\be \label{qh}
Z_H [b,A,\eta^c,\overline {\eta}^c] = \int [d{\cal M}_h]\exp\left\{
iW_{\rm eff}({\cal M}_h)-i(G_{{\cal M}_h}(b,A),\eta^c 
\overline {\eta}^c)\right\},
\ee
where $W_{\rm eff}({\cal M}_h)$ is the effective action for the bilocal
fields  ${\cal M}_h$
\be \label{qbw}
W_{\rm eff}({\cal M})=-\frac{1}{2}
\left({\cal M},{\bf K}^{-1}(b,A){\cal M}\right)
-i ~\mbox{tr} ~\mbox{ln}G_{\cal M}(b,A)~.
\ee
with the non-Abelian Coulomb kernel ${\bf K}(b,A)$ which depends on the
transverse gluon fields $A$ and the $\mbox{tr}$-symbol includes the
trace over color indices.

This effective action contains the $\eta_0$ -meson  part  (~\ref{327})
\be
\label{AB}
W_{\rm eff}^{[0]}(\eta_0)= \int dt\left[
\frac 1 2\left(\dot\eta^2_0 -m_0^2\eta_0^2\right) V +
C_{\eta}\eta_0(t)\frac{d}{dt}X[b+A] \right]~,
\ee
where
$$
C_{\eta}= \frac{N_f}{F_{\eta}}\sqrt{\frac{2}{3}};~~F_{\eta}\sim 1.1 ~F_{\pi}~,
$$
$F_{\pi}=92.4$ MeV is the pion weak decay constant, $N_f$ is the number of 
flavors, and $X[A]$ is the topological winding number functional (\ref{e1}). 
On the level of conventional transverse variables~\cite{sch,al},
the problem appears, how to extract the dynamics of
the collective variable $X[A]$ from the QCD action.
This calculation also meets the set of problems connected with
\begin{enumerate}
\item the Gribov ambiguity~\cite{g} due to the zero mode $D^2(b)=0$ in
the perturbation theory,
\item the violation of gauge invariance due to the anomaly (\ref{AB}),
\item the violation of translational invariance due to the external field
$b({\vec x})$,
\item missing Lorentz covariance.
\end{enumerate}
Let us consider in the following how the introduction of topological Dirac 
variables helps to cure all these problems.

\subsection{Solution by topological Dirac variables}

First of all, to remove the Gribov ambiguity, we should
extract zero modes of the Gauss law constraint
to define the longitudinal function $f$ and the non-Abelian instantaneous
interactions on the class of functions without zero mode
where the Faddev-Popov determinant is not equal to zero.

Then, we construct the topological Dirac variables
\be
A^T+(b^T) = {\bar U}[\Phi] [A +(b+\partial)]{\bar 
U}[\Phi]^{-1};~~~\psi^T={\bar U}[\Phi]\psi
\ee
and announce that these variables correspond to observables.
As a consequence,  the topological functional $X$ converts into
the independent global variable $N_0$ and mixes with the
$\eta_0$-meson channel thus changing its mass.

The effective hadronic Lagrangian in the pseudoscalar, isoscalar
($\eta$-meson) channel in its rest frame has the form \cite{2}
\be \label{327}
L_{eff}(N_0,\eta_0)=
 \frac{{\dot N_0}^2M_N}{2}+
 \frac 1 2({\dot\eta_0}^2-m_0^2\eta_0^2)V
+C_{\eta}\eta_0(t)\dot N_0~.
\ee
After diagonalization it reveals the mass shift of the $\eta_0$-meson
\be   \label{ano}
L_{eff}(\overline N_0,\eta_0)=
  \frac{{\dot {\overline N_0}}^2M_N}{2}+
 \frac 1 2[{\dot\eta_0}^2-(m_0^2+\Delta m^2)\eta_0^2]V~,
\ee
where
\be
{\dot {\overline N_0}}={\dot N_0}+C_{\eta}\eta_0(t)/M_N
\ee
is a new topological variable, and
\be \label{meson}
\Delta m^2=\frac{(C_{\eta})^2}{M_N V}=
\frac{\alpha_s \mu^4 N_f}{6\pi F_{\pi}}
\ee
is the mass shift of the $\eta_0$-meson which is in agreement with the
experimental data when the Shifman-Vainshtein-Zakharov condensate value
(\ref{cd}) $<\alpha_sG^2(b)>=2\mu^4$ \cite{bl} is used.

Thus, we define both the gauge-invariant variables in terms of which
the procedure of hadronization should be fulfilled and the
effective infrared parameter of hadronization $\mu$.

\subsection{Topological confinement}

Another consequence of the topological gauge-fixing of sources of
the observable fields  is the nonperturbative zero-mode phase factor in 
front of any colored state, which depends on the topological variable $N_0$.
In perturbation theory, instead of the colored states in the conventional
gauge $<1|a^+|\psi^D(\vec x)>$, we get the topologically 'dressed' ones
$$
<1|a^+|\psi^T({\vec x}))>=
\exp[N_0{\hat \Phi}_0({\vec x})]<1|a^+|\psi^D({\vec x})>.
$$
Fixing the topological momentum
as a conserved quantum number requires the averaging of colored state 
amplitudes over the parameter of degeneration $N_0$ and over the Euler 
angles of the matrix $\Omega$ in the color space.

This averaging can lead to confinement as a complete destructive
interference phenomenon \cite{vp2,pn}.
Thus, all colored local variables become ghosts. 
They occur in Feynman diagrams but not in the observable asymptotic states. 
The observable states are colorless bound states which depend on the 
translation-invariant relative coordinate $(x-y)$.

\subsection{Relativistic covariance}

In QED, in terms of the Dirac variables, the Poincare
symmetry is realized which is mixing with the gauge one \cite{z}.
This mixing was interpreted in 1930 by Heisenberg and Pauli~\cite{hp} (with
reference to the unpublished note by von Neumann) as the transition from the
Coulomb gauge with respect to the time axis in the rest frame
($l_{\mu}^0=(1,0,0,0)$) to the Coulomb gauge with respect to the
time axis in the moving frame.
$$
l_{\mu}=l_{\mu}^0 + \delta_L {l_{\mu}}^0 \;=\; {(L l^0)}_{\mu}.
$$

The Coulomb interaction has the covariant form
\begin{eqnarray}
{ W}_{C}
= \int d x d y \frac{1}{2}
j_{l}^{T}(x)
V_C(z^{\perp})
j_{l}^{T}(y) \delta(l \cdot z) \,\,\, ,
\end{eqnarray}
where
\begin{eqnarray}
j_{l}^{T} = e \bar {\psi}^T \rlap/l \psi^T \,\, , \,\,
z_{\mu}^{\perp} =
z_{\mu} - l_{\mu}(z \cdot l) \,\, , \,\,
z_\mu = (x - y)_\mu  \,\, . \,\,
\end{eqnarray}
This transformation law of the  Lorentz covariance  was reproduced
in quantum theory by Zumino \cite{z}.

In the non-Abelian theory, in the moving Lorentz frame an observer
sees the moving Wu-Yang monopole \cite{fn} with the corresponding
dynamics.
We should only manage to realize this transformation law on the level
of operators.

\section{Conclusion}

In this paper, we have presented a scheme for the introduction of
topological gauge-invariant variables in QCD by generalizing the 
concept of Dirac variables from QED to non-Abelian gauge theories for the 
example of two-color QCD. 
We construct a model for these variables using a zero mode of the Gauss
law constraint from the class of functions
which disappear at spatial infinity but have nonzero surface integrals.
The model is based on the condition that the winding number functional 
converts into the zero-mode.
It is shown that the corresponding equation of invariance has a solution
for asymptotically stationary fields in the form of the Wu-Yang monopole.

Following Wu and Yang~\cite{wy}, we consider this monopole
except for the singularity at the origin.
In the considered region of the space, the zero-mode
solution of the Gauss law constraint forms
the $\theta$-vacuum \cite{col,jackiw} and leads to a finite
constraint-shell action for definite values of the strong coupling
constant.

To construct the generating functional of the quantum field theory,
it requires only the constraint-shell action, where the (global) 
zero-mode part represents surface integrals defined
in the region at spatial infinity. 
We suggest that in this region the gauge fields are stationary and
the zero-mode solution is factorizing.

In the field of the Wu-Yang monopole the instantaneous quark-quark
potential is a sum of the Coulomb potential and
the golden-section one. The latter one can lead to spontaneous
chiral symmetry breaking and to mesonic bound states. 
The $\eta'$-meson mixes with the zero-mode so that after
diagonalization of this low-energy action a mass shift of the 
$\eta'$-meson is obtained which resolves the $U_A(1)$ problem.

Color amplitudes contain additional phase factors
which depend on the zero-mode.
Averaging  the color amplitudes over the zero-mode parameters
leads to the phenomenon of complete destructive interference
\cite{vp2}, so that the color amplitudes disappear.

According to Heisenberg, Pauli \cite{hp} and Zumino \cite{z}, 
the relativistic covariance is established by a rotation of the timelike 
axis so that the Coulomb field moves together with relativistic bound states.
Recently, Faddeev and Niemi~\cite{fn} constructed a similar 
relativistically covariant form of effective Lagrangian using the 
Wu-Yang monopole.

In summary the present scheme for the introduction of topological 
gauge-invariant variables is a promising tool for the investigation
of the challenging properties of QCD such as the meson spectrum, 
chiral symmetry breaking, quark and gluon confinement, and the $U_A(1)$ 
anomaly. Detailed numerical analyses and the generalization to $SU_c(3)$ 
are to be presented in a subsequent work.

\section*{Acknowledgements}

\medskip

We thank  Profs. A. V. Efremov, G. A. Gogilidze, V. G. Kadyshevsky,
A. M. Khvedelidze, E. A. Kuraev, and L. Lusanna for interesting and critical
discussions. One of us (VNP) is grateful for a stipend from the 
{\sc Max-Planck-Gesellschaft} for his study visit at the University of Rostock.

\end{document}